\documentclass[11pt]{article} %
\usepackage[preprint]{acl}

\usepackage{latexsym}
\usepackage[T1]{fontenc}
\usepackage[utf8]{inputenc}
\usepackage{hyperref}
\usepackage{microtype}
\usepackage{inconsolata}

\usepackage{multirow}
\usepackage{url}
\usepackage{algorithm}
\usepackage{algorithmic}
\usepackage{enumitem}
\usepackage{newfloat}
\usepackage{listings}
\usepackage{fontawesome5}
\usepackage{newfloat}
\usepackage{times}
\usepackage{helvet}
\usepackage{courier} 
\usepackage{graphicx} 
\usepackage{enumerate}
\usepackage{xspace}
\usepackage{amsmath}
\usepackage{amssymb}
\usepackage{booktabs}
\usepackage[most]{tcolorbox}
\usepackage{subcaption} 
\usepackage{tabularx}  %
\usepackage{makecell}  %
\usepackage{wrapfig}
\usepackage{arydshln}

\title{ReCode: Reinforcing Code Generation with Reasoning-Process Rewards}

\author{
Lishui Fan$^{1,2}$\thanks{Equal contribution.},
Yu Zhang$^{1}$\footnotemark[1],
Mouxiang Chen$^{1,2}$, 
Zhongxin Liu$^{1,2,3}$\thanks{Corresponding author.} \\
$^{1}$College of Computer Science and Technology, Zhejiang University\\
$^{2}$The State Key Laboratory of Blockchain and Data Security, Zhejiang University \\
$^{3}$Hangzhou High-Tech Zone (Binjiang) Institute of Blockchain and Data Security \\
}

\newcommand{\app}{{CG-GRPO}\xspace}
\newcommand{\appben}{{LCB-RB}\xspace}
\newcommand{\appre}{{CRPL}\xspace}
\newcommand{\aapp}{{ReCode}\xspace}
\newtcolorbox[auto counter, number within=section]{customprompt}[2][]{%
    colback=gray!10,            %
    colframe=gray!80,           %
    coltitle=black,             %
    fonttitle=\bfseries,        %
    fontupper=\small,
    colbacktitle=gray!30,       %
    title={#2},                 %
    #1                          %
}

\newcommand{\figref}[2]{\ref{#1}(\subref{#2})}

\makeatletter
\renewcommand{\p@subfigure}{\thefigure} 
\makeatother
\begin{document}

\maketitle

\begin{abstract}

In practice, rigorous reasoning is often a key driver of correct code, while Reinforcement Learning (RL) for code generation often neglects optimizing reasoning quality. 
Bringing process-level supervision into RL is appealing, but it faces two challenges. First, training reliable reward models to assess reasoning quality is bottlenecked by the scarcity of fine-grained preference data. Second, naively incorporating such neural rewards may suffer from reward hacking.
This work proposes \aapp (\textbf{Re}asoning-\textbf{Re}inforced \textbf{Code} Generation), a novel RL training framework comprising: (1) Contrastive Reasoning-Process Reward Learning (\appre), which trains a reward model with synthesized optimized and degraded reasoning variants to assess the quality of reasoning process; and (2) Consistency-Gated GRPO (\app), which integrates the reasoning-process reward model into RL by gating neural reasoning-process rewards with strict execution outcomes, using execution correctness as a hard gate to mitigate reward hacking.
Additionally, to assess the reward model’s discriminative capability in assessing reasoning-process quality, we introduce LiveCodeBench-RewardBench (\appben), a new benchmark comprising preference pairs of superior and inferior reasoning processes tailored for code generation.
Experimental results across HumanEval(+), MBPP(+), LiveCodeBench, and BigCodeBench show that a 7B model trained with \aapp outperforms the base version by 16.1\% and reaches performance comparable to GPT-4-Turbo. We further demonstrate the generalizability of \aapp by extending it to the math domain.

\end{abstract}

\section{Introduction}

Reinforcement Learning (RL) has emerged as a transformative paradigm for advancing Large Language Models (LLMs) in code generation~\cite{cao2026qwen3,tang2026execverify,deepcoder2025}. However, existing approaches primarily rely on outcome-based supervision, such as test-case pass rates~\citep{guo2025deepseek,zeng2025acecoder},
which provides limited guidance on how to reach correct programs. In practice, rigorous reasoning is often a key mechanism that supports correct code, rather than merely co-occurring with it~\citep{wei2022chain,lyu2023faithful,fan2025sek}. Consistently, our preliminary investigation shows a significant statistical association between reasoning-process quality and solution correctness (see Appendix~\ref{append:corr}). Therefore, a pivotal question arises: \textit{Can we exploit reasoning-process quality as an additional training signal—beyond final outcome—to improve code generation?}

However, translating this idea into a practical training framework presents two challenges. First, providing scalable, fine-grained supervision over reasoning processes is non-trivial. While utilizing strong LLMs as judges offers a potential solution, their high computational cost and latency render them impractical for the dense sampling required in RL training loops~\cite{kim2023prometheus}. A learned reasoning-process reward model provides a scalable way to supply such supervision for RL, but training it is difficult due to the scarcity of fine-grained preference data over reasoning processes. Specifically, human annotation is unscalable and subjective, while utilizing LLMs-as-a-judge often suffers from poor calibration when assigning scalar scores to nuanced reasoning process qualities~\citep{feng2024numerical,ahn2024large}. Second, even with a learned reward model for reasoning processes, integrating it into RL poses risks. Naively incorporating neural rewards may suffer from reward hacking~\citep{guo2025deepseek}. In code generation, unit-test outcomes are strict and verifiable signals, whereas neural rewards are less constrained, making them easier to exploit.

To tackle these challenges, we introduce a reasoning-reinforced framework in RL for code generation. Specifically, we treat reasoning processes as complementary learning signals and require them to be (i) reliably measured and (ii) safely integrated into RL. This leads to \textbf{Re}asoning-\textbf{Re}inforced \textbf{Code} Generation (\aapp), a novel RL framework with two components, i.e., Contrastive Reasoning-Process Reward Learning (\appre) for reliable measurement of reasoning quality and Consistency-Gated GRPO (\app) for safe integration of reasoning-process rewards in RL.
\appre performs contrastive data synthesis to generate \textit{Optimized} and \textit{Degraded} variants of reasoning processes by perturbing three key reasoning-process features, i.e., factual accuracy, logical rigor, and coherence, and thereby forms a multi-level preference ordering. From this ordering, \appre derives fine-grained preference pairs and trains a reward model for reasoning-process discrimination.
\app augments GRPO by using the functional correctness of generated code as a hard gate for neural reasoning-process rewards, applying it only when the code runs correctly. This gating mechanism enforces reasoning–solution consistency, mitigating reward hacking while preserving informative gradients among correct solutions.

In addition, to assess how well the reward model can discriminate between superior and inferior reasoning processes, we construct LiveCodeBench-RewardBench (\appben), a benchmark comprising 219 manually-checked preference pairs of reasoning processes derived from LiveCodeBench~\citep{jain2025livecodebench}. 
To our knowledge, this is the first benchmark focusing on reasoning-process discrimination in code generation.

Extensive experiments demonstrate the effectiveness of \aapp. On four benchmarks, i.e., LiveCodeBench, HumanEval(+), MBPP(+), and BigCodeBench, Qwen2.5-Coder-7B-Instruct trained with \aapp achieves a relative improvement of 16.1\% over the base model (50.4\%$\rightarrow$58.5\%) and
exhibiting performance comparable to GPT-4-Turbo. Additional results show that \aapp generalizes to Qwen3-4B and complements compiler-based process supervision. Meanwhile, reward models trained with \appre achieve strong discriminative performance on \appben and also perform well on the reasoning subset of RewardBench~\cite{lambert2024rewardbench}, indicating that the model trained with \appre learns a reasoning-process reward that generalizes and is consistent with functional correctness. We further observe consistent gains by extending \app to the math domain.  The models, datasets, and code are publicly available\footnote{https://github.com/ZJU-CTAG/ReCode}.

\begin{figure*}[htbp]
    \centering
    \includegraphics[width=0.9\linewidth]{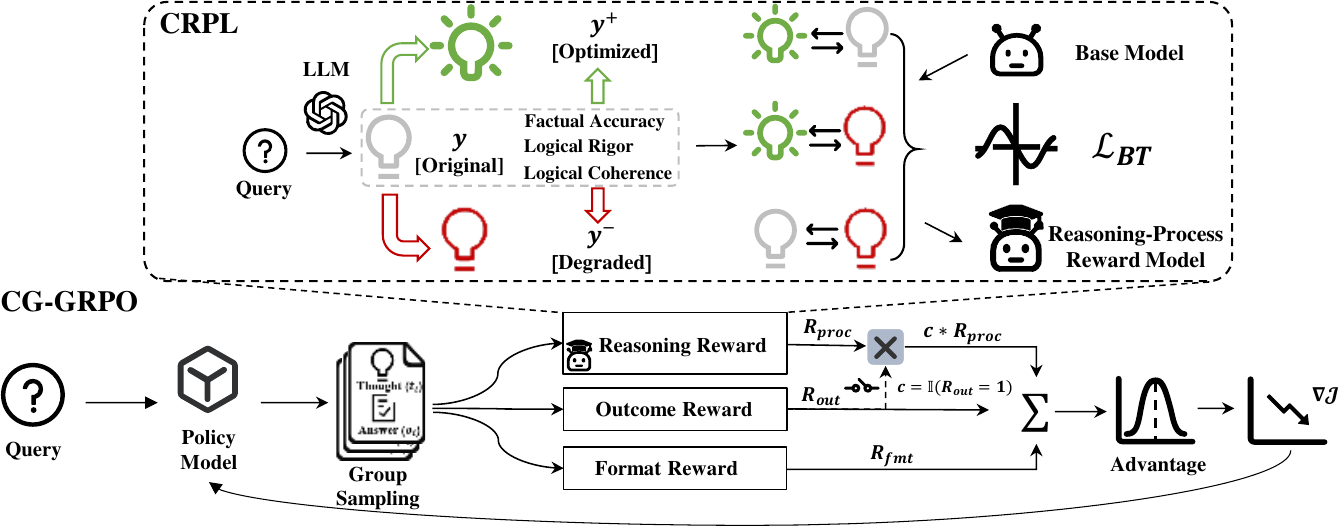}
    \caption{Overview of \aapp. \textbf{Top}: \appre for reliably measuring reasoning quality. \textbf{Bottom}: \app for safely integrating reasoning-process rewards into RL.} 
    \label{fig:main}
\end{figure*}
\section{\aapp}
\label{sec:method}

\subsection{Contrastive Reasoning-Process Reward Learning}
\label{subsec:reward_model}
The effectiveness of incorporating reasoning-process rewards into RL hinges on a reward model that can provide a reliable measurement of reasoning quality. However, training such a model is bottlenecked by the scarcity of fine-grained preference data over reasoning processes.
To address this, we propose Contrastive Reasoning-Process Reward Learning (\appre) to automatically synthesise contrastive preference data by generating optimized and degraded reasoning variants, and train a reasoning-process reward model on the resulting preference data. As shown in Figure~\ref{fig:main}, \appre induces a multi-level preference ordering over reasoning processes and trains the reward model to discriminate reasoning-process quality. The reward model then produces reasoning-process rewards for the policy optimization module.

To construct high-quality, reasoning-aware preference data, we first codify the intrinsic attributes of a reasoning process. Drawing upon insights from manual analysis and prior work~\cite{zhu2025chain}, we identify three critical dimensions of reasoning quality: (1) \textit{Factual Accuracy}: assesses whether the reasoning contains factual errors; (2) \textit{Logical Rigor}: assesses (a) whether redundant or misleading logical steps exist, and (b) whether missing logical connections result in incomplete reasoning; (3) \textit{Logical Coherence}: assesses whether the logical flow maintains clear connections between steps. 

We then prompt a powerful LLM, Qwen2.5-Coder-32B-Instruct, to synthesize a base reasoning process $t$ for a given problem $x$, and leverage the dimensions mentioned above to generate variants of the base reasoning process $t$. Instead of generic rewriting, which may yield stylistic changes or ambiguous quality differences, we apply targeted transformations along specific reasoning-process dimensions to construct a multi-level preference ordering, providing clearer preference supervision for reward modeling.
Concretely, we generate both optimized and degraded variants. (a) For optimized variants ($t^+$), the model is prompted to enhance specific dimensions, such as removing redundancies to improve logical rigor or clarifying transitions to boost logical coherence. (b) For degraded variants ($t^-$), the model is prompted to inject subtle flaws based on these dimensions, such as introducing a factual error or a logical gap. The optimization and degradation prompts are in Appendix~\ref{append:prompt}.

We formulate three distinct types of preference pairs to establish a continuous quality manifold: 
(a) \textit{strong contrast pairs} $(x, t^+ \succ t^-)$,
(b) \textit{fine-grained optimization pairs} $(x, t^+ \succ t)$, and 
(c) \textit{fine-grained degradation pairs} $(x, t \succ t^-)$.
This comprehensive training paradigm provides comparative signals that enable the reward model to internalize the intrinsic features of high-quality reasoning through relative contrast rather than absolute quantification. 
Notably, while similar evolutionary strategies like Evol-Instruct~\citep{xu2024wizardlm} have been explored for synthesizing supervised fine-tuning data,  we novelly identify and validate their effectiveness in training discriminative reward models for distinguishing nuances in reasoning-process quality.

\textbf{Reward modeling objective.} Let $s_{\theta}(x,t)\in \mathbb{R}$ be the score given by the reward model $s_{\theta}$ for the reasoning process $t$ under problem $x$. For each preference pair $(x,t^{(a)} \succ t^{(b)}) \in \mathcal{D}$, where $t^{(a)}$ is preferred over $t^{(b)}$,
we optimize a Bradley–Terry~\citep{bradley1952rank} objective, which is widely adopted in reward modeling~\citep{ouyang2022training}:
$$
\mathcal{L}_{BT}(\theta) = -\log \sigma( s_{\theta}(x, t^{(a)}) - s_{\theta}(x, t^{(b)}))
$$
and sum over all constructed pairs. 
Minimizing this objective encourages $s_{\theta}$ to assign higher scores to higher-quality reasoning processes.

\subsection{Consistency-Gated GRPO}
\label{subsec:RL}
Given the \appre-trained reasoning-process reward model, this section instantiates the policy-optimization module that jointly leverages functional correctness and reasoning-process signals. 
Our implementation builds upon GRPO~\citep{shao2024deepseekmath}, which uses group-relative baselines computed from a set of sampled responses to reduce variance. In our setting, each response $y$ is explicitly decomposed into a reasoning process and a solution,
$y \triangleq \big(t, o\big)$
, where $t$ denotes the content inside the \texttt{<think>} block and $o$ denotes the content inside the \texttt{<answer>} block. 

While GRPO can optimize binary outcome rewards effectively, naively incorporating a neural reward makes the policy susceptible to reward hacking~\citep{guo2025deepseek}. For instance, the policy may learn to inflate the process score without improving the functional correctness of $o$. To mitigate this, we propose Consistency-Gated GRPO (\app), a simple and effective gating mechanism that makes process optimization conditional on successful execution. 
As shown in Figure~\ref{fig:main}(c), \app introduces a reward structure composed of three terms:

\noindent \textit{(1) Format Reward ($R_{fmt}$):} A binary reward ensuring structural compliance (i.e., correct usage of \texttt{<think>} and \texttt{<answer>} tags). $R_{fmt} = \mathbb{I}(\text{Valid Format})$, which has been demonstrated to be effective in prior works~\citep{xie2025logic,guo2025deepseek}.
 \textit{(2) Outcome Reward ($R_{out}$):} A strict binary signal derived from test case execution. $R_{out} = 1$ if the solution passes all test cases, and $0$ otherwise.
 \textit{(3) Reasoning Reward ($R_{proc}$):} A continuous scalar $r \in [0, 1]$ provided by the \appre-trained reward model $s_{\theta}$ (Section~\ref{subsec:reward_model}), quantifying the quality of the reasoning process $t$.

A naive way to incorporate $R_{proc}$ is to linearly combine all reward terms (e.g., $R = R_{fmt}+R_{\text{out}} + R_{\text{proc}}$). 
However, since $R_{\text{proc}}$ is a neural signal and less constrained than the signal provided by execution feedback, such linear aggregation renders the policy susceptible to reward hacking (see Section~\ref{sec:discussion}), where the policy inflates process scores without improving functional correctness.  
To mitigate this, our consistency gate grounds $R_{\text{proc}}$ in a strict and verifiable signal. We activate $R_{\text{proc}}$ only for functionally correct samples. In this way, functional correctness acts as a hard constraint preventing the policy from compromising correctness for higher neural scores, while still providing informative gradients to distinguish reasoning processes among correct solutions. The final reward $R_i$ for the $i$-th sample is formulated as:

$$
    R_i = 
    \underbrace{R_{fmt}^{(i)} \vphantom{\mathbb{I}(R_{out}^{(i)}=1)}}_{\text{Structure}} + 
    \underbrace{R_{out}^{(i)} \vphantom{\mathbb{I}(R_{out}^{(i)}=1)}}_{\text{Correctness}} + 
    \underbrace{\mathbb{I}(R_{out}^{(i)}=1) \cdot R_{proc}^{(i)}}_{\text{Consistency-Gated Process}}
$$
where $\mathbb{I}(\cdot)$ is the indicator function. This formulation ensures that the process reward is activated if and only if the extrinsic outcome is correct. 

A key advantage of \app is its ability to maintain learning dynamics when the model performs well. In standard GRPO, if a group of $G$ samples is all correct ($R_{out}^{(i)}=1, \forall i \in G$), the standard reward is uniform, leading to zero advantage. 
In contrast, our gated term $\mathbb{I}(R_{out}=1) \cdot R_{proc}$ introduces meaningful variance among functionally correct solutions when their reasoning-process quality differs. This creates non-zero advantages, encouraging the model to prefer high-quality reasoning processes among correct solutions.

\section{Benchmark Construction}

\label{subsec:benchmark}
Existing benchmarks for evaluating reward models~\citep{lambert2024rewardbench,liu2025rm} primarily focus on distinguishing the correctness of final solutions rather than the quality of the intermediate reasoning process. This outcome-centric focus renders these benchmarks insufficient for evaluating reward models for training processes. To bridge this gap, we introduce LiveCodeBench-RewardBench (\appben), a benchmark designed to discriminate between superior and inferior reasoning processes.

Following established protocols~\citep{lambert2024rewardbench,liu2025rm}, we use Qwen2.5-Coder-32B-Instruct with high-temperature sampling ($T=1.0$) to generate 50 reasoning-solution pairs per problem from LiveCodeBench v6.
We first partition the generated solutions into pass and fail sets based on test-case execution.
In most cases, execution outcomes provide a useful coarse signal of reasoning-process quality-correct solutions are more likely to be supported by coherent reasoning than incorrect ones.
However, our manual analysis reveals a small but important \emph{Reasoning-Implementation Gap}:
(i) a solution with a high-quality reasoning process may fail due to minor implementation issues (e.g., missing imports), and
(ii) a solution with a flawed reasoning process may spuriously pass test cases due to randomness~\citep{wang2023cost}.

To purify the data and ensure strict alignment between reasoning and implementation, we implement a two-stage filtration pipeline:

\textit{Stage 1: Automated Dual-Consistency Check.}
We employ GPT-4o~\citep{Achiam2023GPT4TR} as an external validator to assess two criteria:
(1) \textit{Logical Soundness}, i.e., whether the reasoning processes contains substantive logical flaws; and
(2) \textit{Implementation Alignment}, i.e., whether the produced code faithfully implements the described plan.
These checks target the two failure modes of the Reasoning-Implementation Gap. The prompt is in Appendix~\ref{append:prompt}.
Based on GPT-4o's assessment of criterion (2), we discard instances with poor alignment (i.e., code that does not implement the stated reasoning).  
Among the remaining aligned instances, we keep (a) logically sound processes judged by GPT-4o as \emph{chosen candidates} and (b) logically flawed processes judged by GPT-4o as \emph{rejected candidates}.
    
\textit{Stage 2: Human Adjudication.}
To mitigate potential bias of LLM-as-a-Judge~\citep{zheng2023judging}, two authors independently inspect the filtered instances. Adhering to the same evaluation criteria and prompt definitions used in Stage 1, the annotators manually evaluate each instance for both logical soundness and implementation alignment.
Inter-annotator agreement measured by Cohen's Kappa is $\kappa=0.769$, indicating substantial agreement~\citep{landis1977measurement}.
Disagreements are subsequently resolved through consensus discussion between the annotators to ensure the high quality of the final labels.

From the adjudicated pool, we construct problem-wise preference pairs.
We first exclude problems lacking either a valid chosen or a valid rejected process to ensure pairwise comparability.
Within the valid problems, we observe that valid rejected processes (clear logical flaws under reliable reasoning-code alignment) remain relatively scarce.
To balance pairs, for each validated rejected process, we randomly sample one validated chosen process from the same problem instance.

We reserve problems from Oct.\ 2024 to Feb.\ 2025 exclusively for final code-generation evaluation to prevent data leakage. Finally, we obtain 219 preference pairs.

\section{Experimental Setup}

\textbf{Reward Model Setup.} Our reward models are initialized from Qwen2.5-Coder-7B-Base and Qwen2.5-Coder-3B-Base, and trained on preference pairs from the DeepCoder-Preview-Dataset~\citep{deepcoder2025}, a corpus of 24k coding problems. We evaluate our reward model on \appben and the code and math subsets of RewardBench~\citep{lambert2024rewardbench}, using accuracy as the evaluation metric~\citep{lambert2024rewardbench,liu2025rm}. Baselines include: (a) the original model, (b) state-of-the-art (SOTA) reward models, including Starling-RM-34B~\citep{starling2023}, EURUS-RM-7B~\citep{yuan2024advancing}, Skywork-Reward-Llama-3.1-8B~\citep{liu2024skywork}, GPT-4-Turbo-2024-04-09~\citep{Achiam2023GPT4TR}, GPT-3.5-Turbo-0125~\citep{brown2020language} and DeepSeek-V3~\cite{liu2024deepseek}, and (c) a Score-Based reward model. Further details are provided in appendix~\ref{append:reward}.

\noindent \textbf{RL Setup.} We select Qwen2.5-Coder-7B-Instruct as policy model, using DeepCoder-Preview-Dataset for training. 
The prompt used during training is in Appendix~\ref{append:promptrl}. We conduct evaluations on HumanEval(+)~\citep{liu2024your,chen2021evaluating}, MBPP(+)~\citep{austin2021program,liu2024your}, BigCodeBench~\citep{zhuo2024bigcodebench}, and LiveCodeBench~\citep{jain2025livecodebench}. We use greedy decoding and employ Pass@1 for evaluation. Baselines include: (a) the original model, (b) SOTA code models, including Llama3-Instruct-70B~\citep{dubey2024llama}, Deepseek-Coder-V2-Lite-Instruct~\citep{zhu2024deepseek}, Qwen2.5-Coder-Instruct 14B~\citep{hui2024qwen2}, GPT-4-Turbo-2024-04-09, and GPT-3.5-Turbo-0125, (c) the model fine-tuned with SFT on the same data, and (d) the model only with outcome and format rewards. Further details provided in Appendix~\ref{append:rl}.

\section{Results}
We aim to answer the following research questions:\\
\noindent \textbf{RQ1}: How effective is \aapp in improving code generation across different benchmarks? \\
\noindent \textbf{RQ2}: How effectively does the \appre-trained reward model distinguish reasoning-process quality on \appben? Does this discriminative capability generalize to other reasoning benchmarks?\\
\noindent \textbf{RQ3}: Can the \aapp training paradigm generalize to mathematical tasks which also rely on high-quality reasoning capabilities?

\subsection{RQ1: Effectiveness of \aapp in Code Generation}
\begin{table*}[t]
\centering
\scriptsize
\resizebox{\linewidth}{!}{%
\begin{tabular}{@{}lccccccccccc@{}}
\toprule
\textbf{Model}                     & \textbf{Size}                 & \multicolumn{2}{c}{\textbf{Humaneval}} & \multicolumn{2}{c}{\textbf{MBPP}}      & \multicolumn{3}{c}{\textbf{LiveCodeBench}}            & \multicolumn{2}{c}{\textbf{BigCodeBench}} & \textbf{Avg}       \\ 
                          & - & \textbf{HE}            & \textbf{HE+}           & \textbf{MBPP}          & \textbf{MBPP+}         & \textbf{Easy}          & \textbf{Medium}        & \textbf{Hard}         & \textbf{Full}            & \textbf{Hard}           & -             \\ \midrule
GPT-4-Turbo               & \faLock & 90.2          & 86.0          & 85.7          & 73.3          & 68.5          & 24.2          & 4.6          & 58.2            & 35.1           & 58.4          \\
GPT-3.5-Turbo             &   \faLock                   & 72.6          & 67.7          & 84.1          & 71.2          & 46.3          & 9.4           & 5.6          & 50.6            & 21.6           & 47.7          \\ \midrule
Qwen2.5-Coder-Instruct    & 14B                  & 89.6          & 87.2          & 86.2          & 72.8          & 61.0          & 11.3          & 2.8          & 48.4            & 22.2           & 53.5          \\
DS-Coder-V2-Lite-Instruct & 2.4/16B              & 81.1          & 75.6          & 82.8          & 70.4          & 43.9          & 5.7           & 5.6          & 36.8            & 16.2           & 46.5          \\
Llama3-Instruct           & 70B                  & 77.4          & 72.0          & 82.3          & 69.0          & 43.9          & 7.5           & 5.6          & 54.5            & 27             & 48.8          \\ \midrule
Qwen2.5-Coder-Instruct    & 7B                   & {88.4} & {84.1} & 83.5          & 71.7          & 56.1          & 3.8           & {6.9}          & 41.0            & 18.2           & 50.4          \\
+SFT                     &   7B                   & 66.2          & 57.3          & 73.3          & 63.5          & 34.1          & 3.8           & 0.0          & 39.9            & 13.5           & 39.1          \\
+GRPO               & 7B                   & 85.9          & {81.1}          & {86.7}          & {75.1}          & {58.5} & {15.1}          & \textbf{9.7}          & {52.0}            & {29.7}           & {54.9}          \\
+\aapp         & 7B                   & \textbf{90.9}          & \textbf{86.0}          & \textbf{87.0} & \textbf{76.2} & \textbf{68.3} & \textbf{20.8} & \textbf{9.7} & \textbf{54.0}   & \textbf{33.8}  & \textbf{58.5} \\ \bottomrule
\end{tabular}%
}
\caption{Performance comparison of Qwen2.5-Coder-Instruct with \aapp against other baselines.}
\label{tab:code-main}
\end{table*}

\begin{figure}[t]
    \centering
    \includegraphics[width=0.9\linewidth]{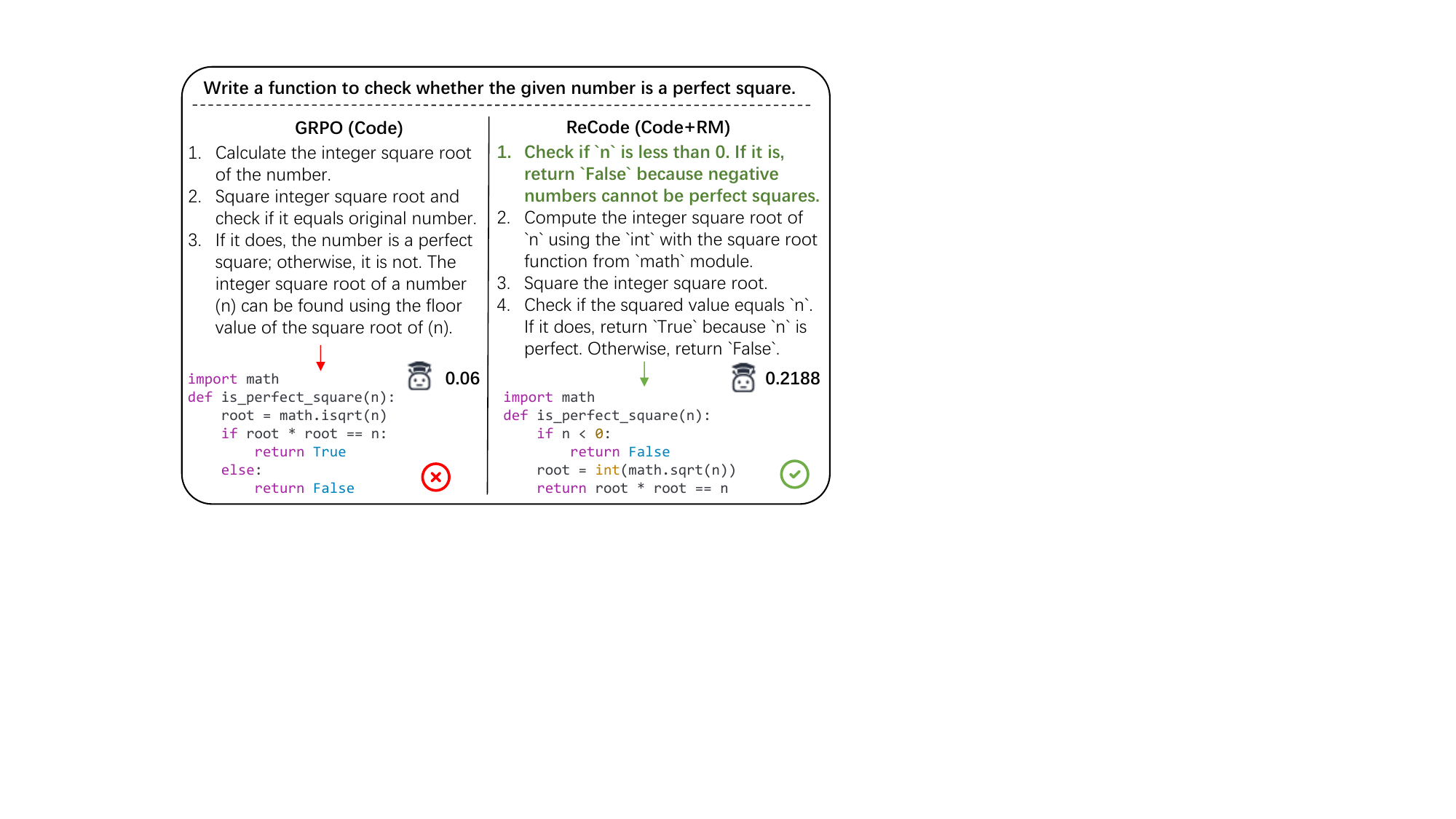}
    \caption{Example of reasoning processes generated by the base model with \aapp and with GRPO.}
    \label{fig:case-mbpp}
\end{figure}
\aapp effectively enhances the performance of the model. As shown in Table~\ref{tab:code-main}, \aapp achieves a relative improvement of 16.1\% on average over the base model across all benchmarks, showing comparable performance to GPT-4-Turbo. Additionally, \aapp surpasses the outcome-only baseline by 6.7\%, maintaining a consistent performance advantage as shown in Figure~\figref{fig:combined-main-results}{fig:lcb-step}.

To further elucidate the mechanisms underlying \aapp's performance gains, we inspect models' outputs. Our analysis reveals that \aapp's primary advantage lies in its ability to generate more comprehensive and logically sound reasoning processes, which help the model produce more accurate code. For example, as shown in Figure~\ref{fig:case-mbpp}, when solving a perfect square problem, the model without a reasoning-process reward fails to consider negative numbers as edge cases in its reasoning process, causing it to pass only the basic test cases while failing the test cases in MBPP+. In contrast, the model with \aapp demonstrates sound reasoning by considering negative inputs at the outset of its reasoning process.

We further investigate the reward score of the generated reasoning process of the case. The results show that the model trained with \aapp achieves a reasoning-process reward of 0.2188, outperforming the model trained with outcome-only rewards, which obtains a score of 0.06. This demonstrates that our reward model is well-calibrated to reasoning-process quality, thereby providing effective supervision to guide the policy optimization. Additional examples can be found in the supplementary materials.

\subsection{RQ2: Reward Model Effectiveness}

\begin{table}[t]
\centering
\scriptsize
\resizebox{\linewidth}{!}{%
\begin{tabular}{@{}lccccc@{}}
\toprule
\textbf{Model} & \textbf{Size} & \textbf{LCB-RB} & \multicolumn{2}{c}{\textbf{RewardBench}} & \textbf{Avg} \\
  & - & - & \textbf{Code} & \textbf{Math} & - \\ \midrule
DeepSeek-V3 & 671B &  \textbf{66.9} &  \textbf{98.5} & 78.5 & \underline{81.3} \\
GPT-4-Turbo & \faLock & \underline{63.7} & \underline{98.1} & 67.3 & 76.4 \\
GPT-3.5-Turbo & \faLock & 50.7 & 77.6 & 40.6 & 56.3 \\ \midrule
Starling-RM & 34B & 55.3 & 88.8 & 85.9 & 76.7 \\
EURUS-RM & 7B & 57.0 & 92.8 & 79.9 & 76.5 \\
\begin{tabular}[c]{@{}l@{}}Skywork\\ -Reward\\ -Llama-3.1\end{tabular} & 8B & 61.6 & - & - & - \\ \midrule
Qwen2.5-Coder & 3B & 50.3 & 52.8 & 60.0 & 54.4 \\
+Score & 3B & 51.0 & 49.4 & 47.2 & 49.2 \\
+\appre & 3B & 57.9 & 63.6 & \underline{93.5} & 71.7 \\ \hdashline
Qwen2.5-Coder & 7B & 53.8 & 43.9 & 65.8 & 54.5 \\
+Score & 7B & 57.7 & 80.2 & 71.8 & 69.9 \\
+\appre & 7B & 62.6 & 88.6 & \textbf{99.8} & \textbf{83.7} \\ \bottomrule
\end{tabular}%
}
\caption{Performance comparison of reward model trained with \appre against other baselines.}
\label{tab:rm-main}
\end{table}

Table~\ref{tab:rm-main} presents the performance of the reward model trained with \appre and other baselines on \appben and the reasoning subsets of RewardBench.
Due to potential data contamination, we exclude the results of Skywork-Reward-Llama-3.1 from RewardBench\footnote{https://gist.github.com/natolambert/\\1aed306000c13e0e8c5bc17c1a5dd300}.
The reward model trained with \appre effectively enhances the base model's ability to identify high-quality reasoning processes on \appben, surpassing other baselines. For instance, our 7B parameter model achieves performance comparable to GPT-4-Turbo.

Compared with the score-based baseline, the \appre-trained models demonstrate substantial improvements. Specifically, our 3B and 7B models achieve relative improvements of 45.7\% and 19.7\% over score-based baselines, respectively. This indicates that training LLMs to distinguish between optimized and degraded versions of reasoning processes is more effective than learning from direct numerical scores. This is likely because LLMs are not inherently sensitive to fine-grained numerical values~\citep{feng2024numerical,ahn2024large}, making it difficult to express the nuanced differences between reasoning processes via a scalar score.

Furthermore, the base model trained with \appre achieves the highest average performance across reasoning subsets of RewardBench, outperforming the best baseline by 6.4\% relatively. As RewardBench evaluates a model’s ability to discriminate the quality of final solutions, this result suggests that the discriminative patterns learned from reasoning processes generalize effectively to outcome assessment.

\subsection{RQ3: Generalization to Mathematical Tasks}

\begin{table}[t]
\centering
\scriptsize
\resizebox{\linewidth}{!}{%
\begin{tabularx}{\linewidth}{@{}l c c c c c @{}}
\toprule
\textbf{Model} & \textbf{Size} & \textbf{\begin{tabular}[c]{@{}c@{}}MATH\\ 500\end{tabular}} & \textbf{\begin{tabular}[c]{@{}c@{}}Minerva \\ Math\end{tabular}} & \textbf{AIME24} & \textbf{Avg} \\ \midrule
GPT-4o & \faLock & 76.4 & 36.8 & 9.3 & 40.8 \\ \midrule
Llama-3.1-Inst & 70B & 64.6 & 35.3 & 16.7 & 38.9 \\
Llama-3.1-Inst & 405B & 73.8 & \textbf{54.0} & 20.0 & 49.3 \\
Eurus-2-PRIME & 7B & 79.2 & 38.6 & 26.7 & 48.2 \\
Qwen2.5-Math-Inst & 7B & 79.8 & 37.1 & 13.3 & 43.4 \\ \midrule
Qwen2.5-Math & 7B & 46.9 & 15.5 & 11.2 & 24.5 \\
+GRPO   & 7B & \textbf{83.0} & 34.2 & 26.7 & 48.0 \\
+\aapp  & 7B & \textbf{83.0} & 38.2 & \textbf{33.3} & \textbf{51.5} \\ \bottomrule
\end{tabularx}%
}
\caption{Performance comparison of Qwen2.5-Math with \aapp against other baselines.}
\label{tab:math-main}
\end{table}

To further assess the generalization of \aapp, we extend it to the math domain where performance hinges critically on high-quality reasoning.

\paragraph{Experimental Setup}
We employ the same reward model in RQ1. For the policy model, we utilize Qwen2.5-Math-7B~\citep{yang2024qwen2}.
We choose DAPO-Math-17k~\citep{yu2025dapoopensourcellmreinforcement} as the training dataset, consisting of 17K mathematical data. We maintain the same experimental setup as in RQ1, except the training steps are reduced to 900, accounting for the smaller dataset size. Model performance is evaluated on MATH500~\citep{hendrycks2021measuring}, Minerva Math~\citep{lewkowycz2022solving} and AIME 2024~\citep{AoPS:AIMEProblemsSolutions}. Following prior work~\citep{cui2025process,dubey2024llama}, we employ greedy decoding for evaluation and report accuracy metric. Baselines include: (1) original model, (2) base model trained via RL without reasoning-process rewards, and (3) SOTA mathematical models, including Llama-3.1-Instruct, GPT-4o-2024-0806, Eurus-2-PRIME~\citep{cui2025process}, 

As shown in Table~\ref{tab:math-main}, \aapp effectively enhances the model's performance on mathematical tasks, indicating that \aapp generalizes beyond code generation. Specifically, the base model with \aapp demonstrates superior performance compared to several SOTA mathematical models. It demonstrates a 7.4\% relative improvement over the RL baseline trained without reasoning-process rewards. To further illustrate the superiority of \aapp, we analyze the performance trajectory on AIME24. As shown in Figure~\figref{fig:combined-main-results}{fig:aime-main}, \aapp outperforms the baseline without reasoning-process rewards throughout the training process.

\section{Discussion}
\begin{figure}[t]
    \centering
    \begin{subfigure}[t]{0.49\linewidth}
        \centering
        \includegraphics[width=\linewidth]{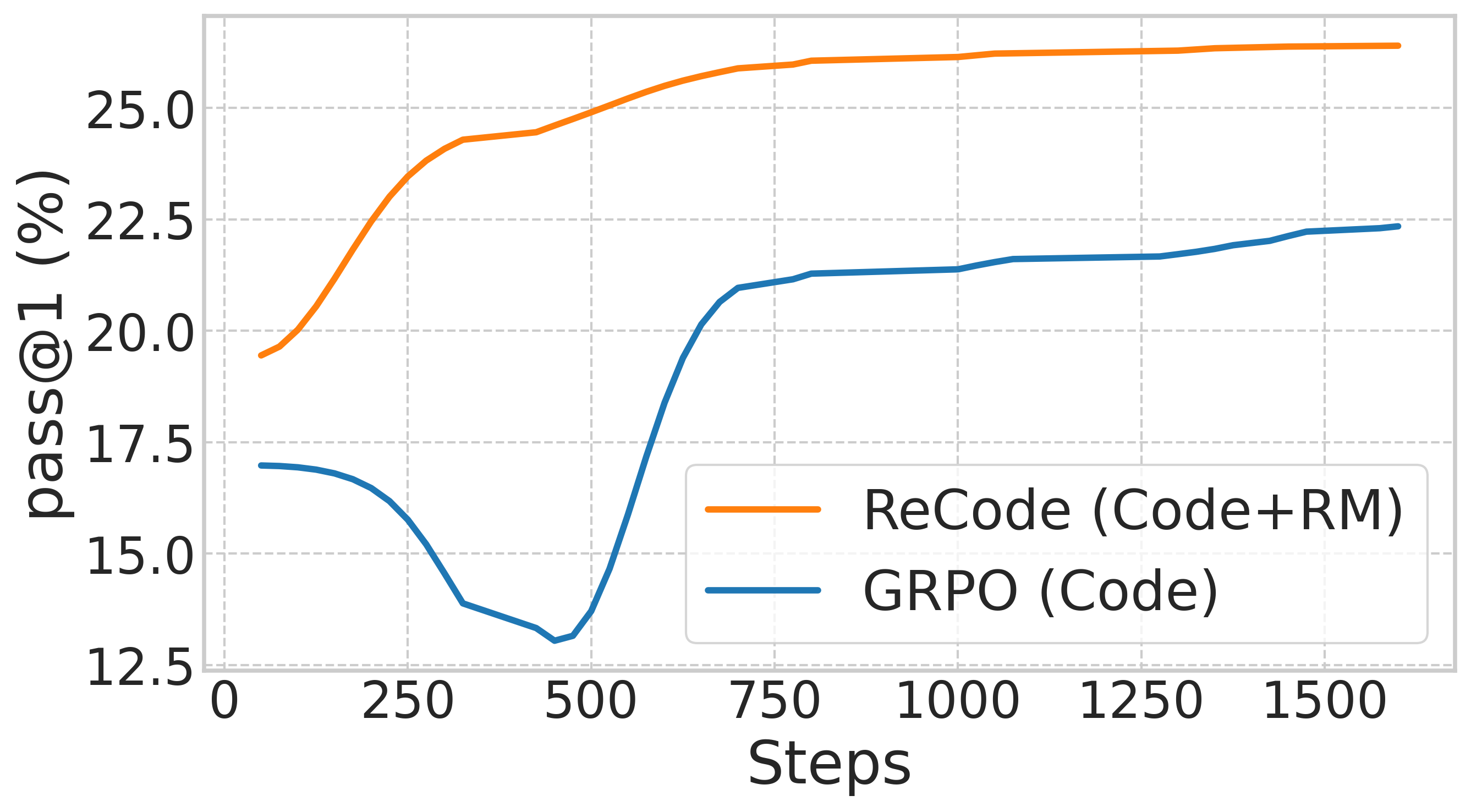}
        \caption{LiveCodeBench.}
        \label{fig:lcb-step}
    \end{subfigure}
    \hfill
    \begin{subfigure}[t]{0.49\linewidth}
        \centering
        \includegraphics[width=\linewidth]{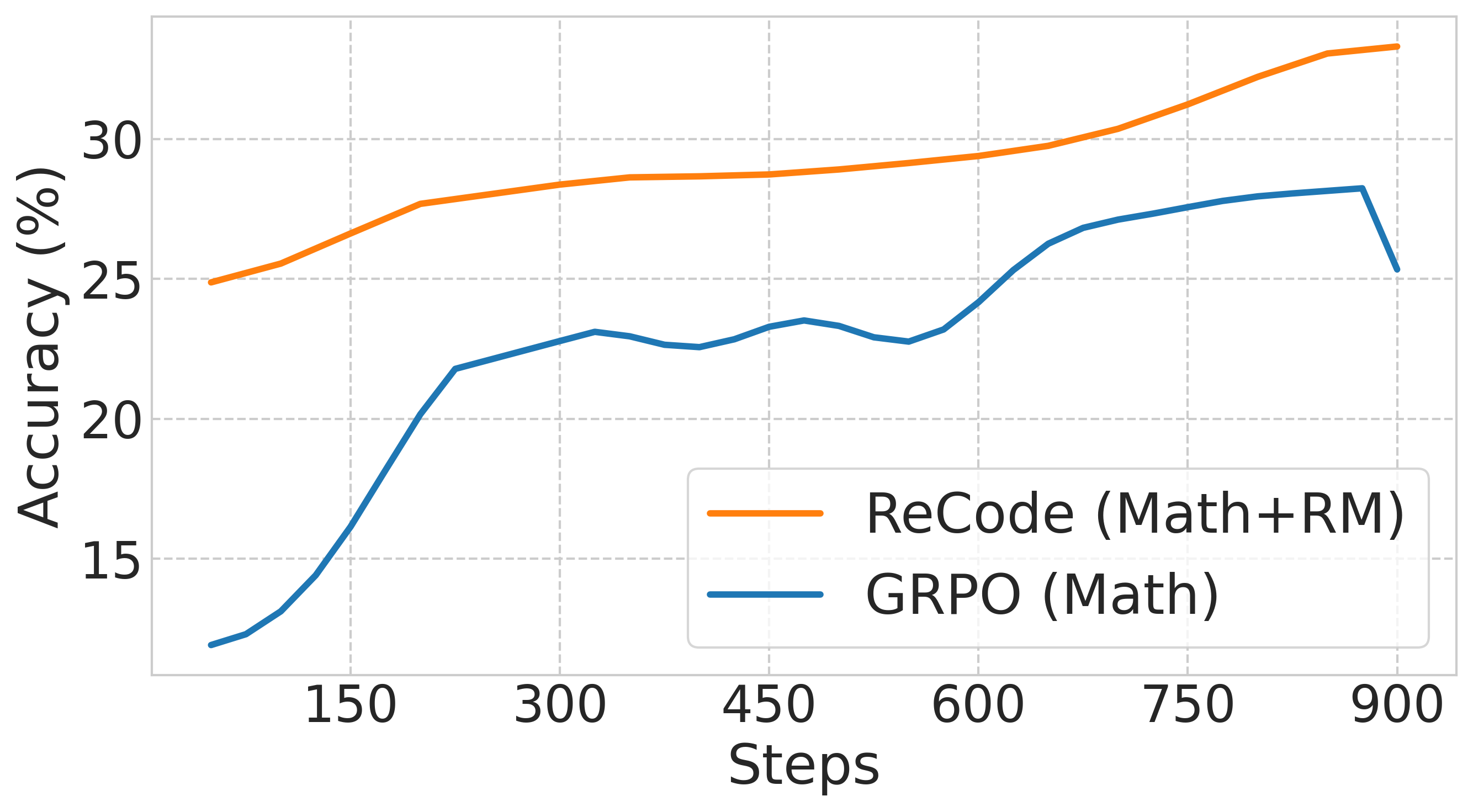}
        \caption{AIME 2024.}
        \label{fig:aime-main}
    \end{subfigure}
    
    \begin{subfigure}[t]{0.49\linewidth}
        \centering
        \includegraphics[width=\linewidth]{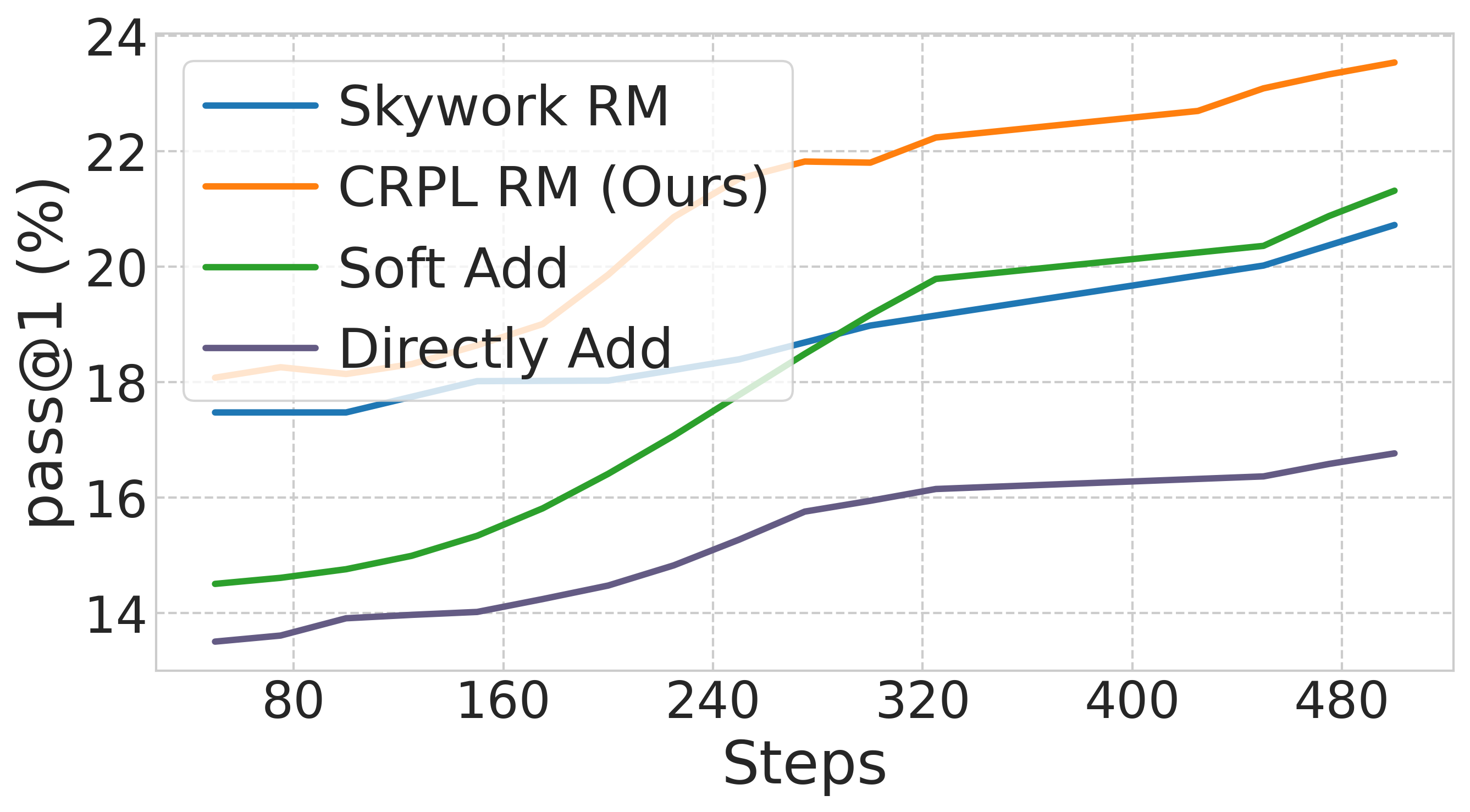}
        \caption{Reward-model and formulation ablations.}
        \label{fig:reward_comparison}
    \end{subfigure}
    \hfill
    \begin{subfigure}[t]{0.49\linewidth}
        \centering
        \includegraphics[width=\linewidth]{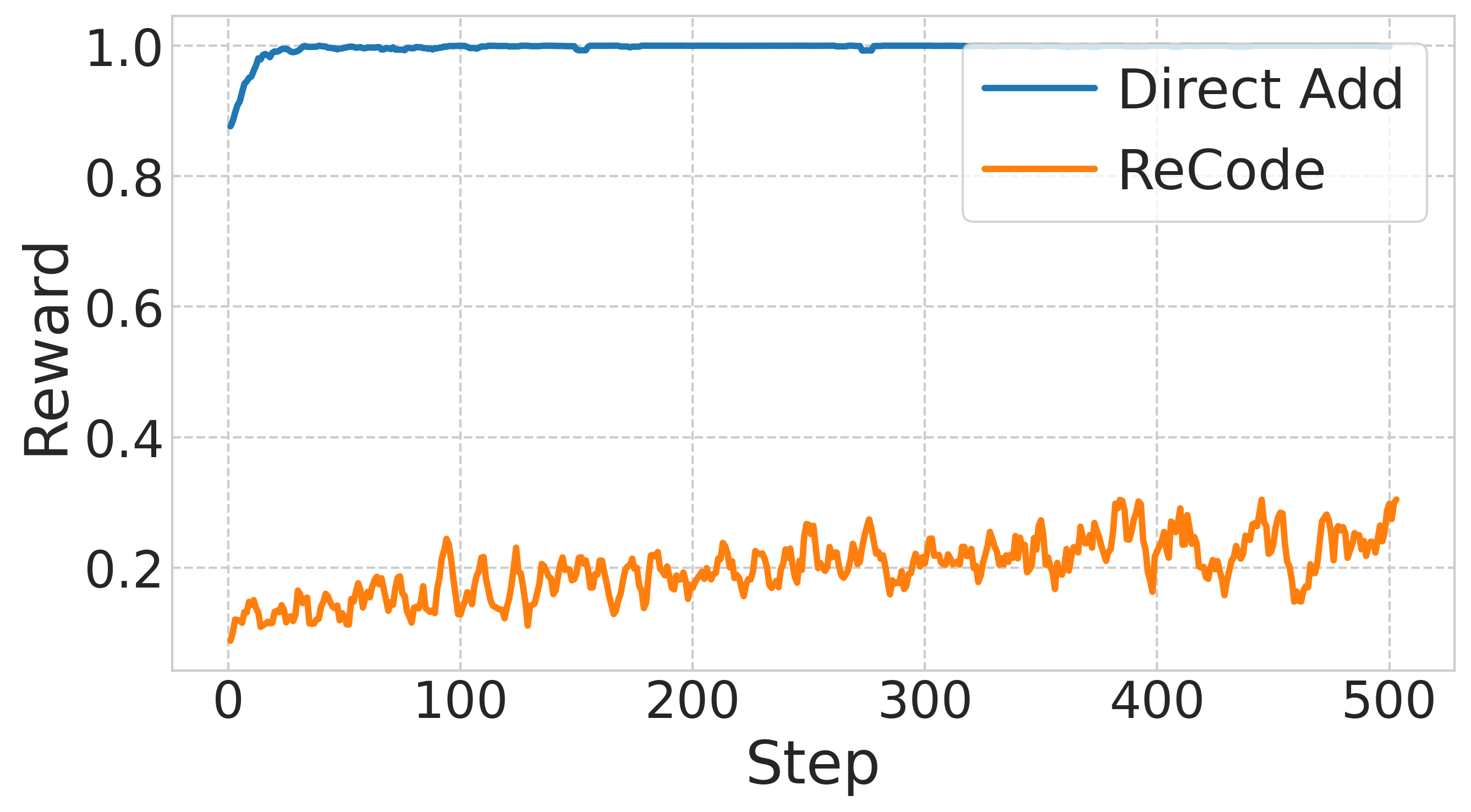}
        \caption{Reward hacking dynamics.}
        \label{fig:reward_hack}
    \end{subfigure}
    \caption{Overall performance of \aapp versus GRPO (a,b), with analyses of reward design (c) and reward hacking (d).}
    \label{fig:combined-main-results}
\end{figure}

\label{sec:discussion}
\begin{figure}[t]
    \centering
    \begin{subfigure}[t]{0.49\linewidth}
        \centering
        \includegraphics[width=\linewidth]{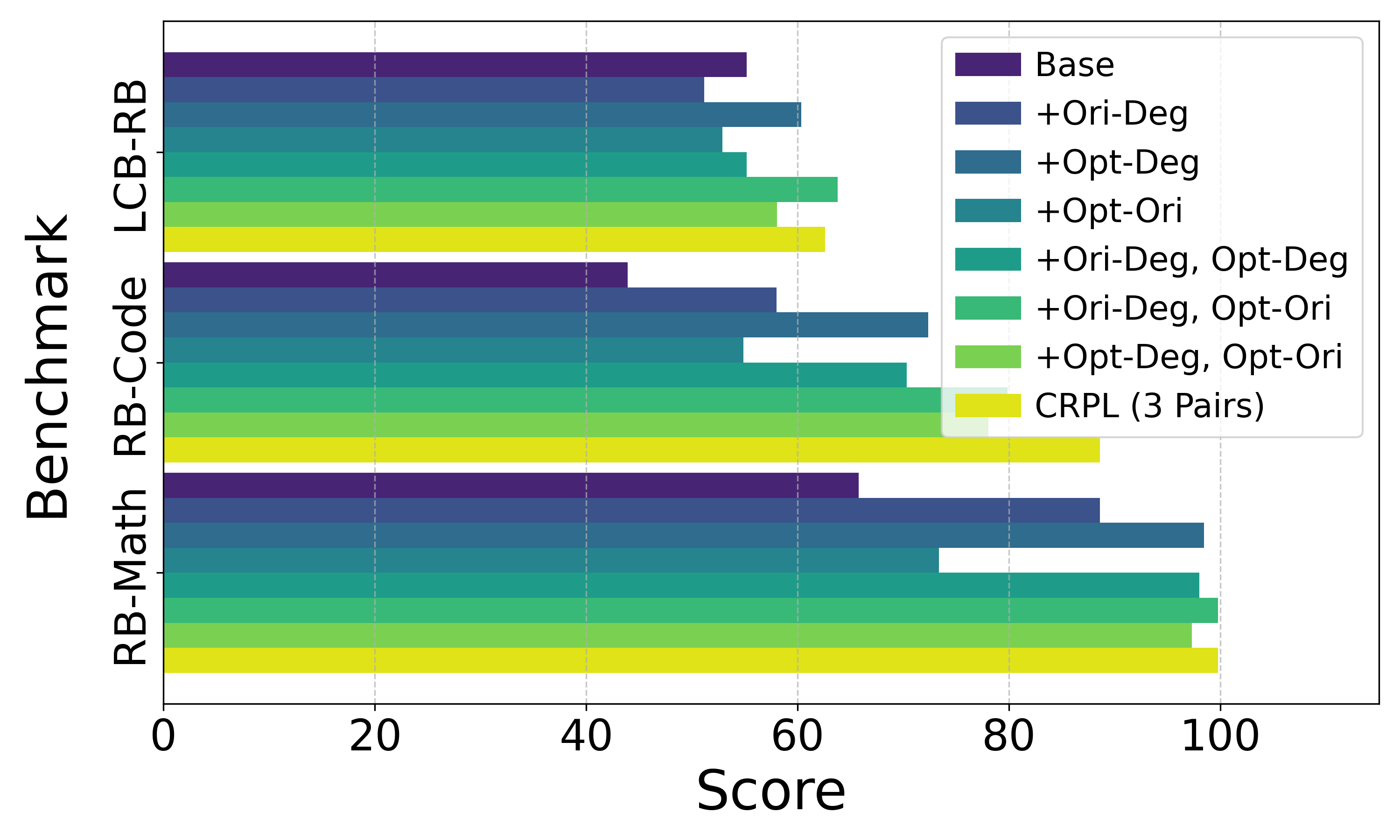}
        \caption{Qwen2.5-Coder-7B.}
        \label{fig:results_7b}
    \end{subfigure}
    \hfill
    \begin{subfigure}[t]{0.49\linewidth}
        \centering
        \includegraphics[width=\linewidth]{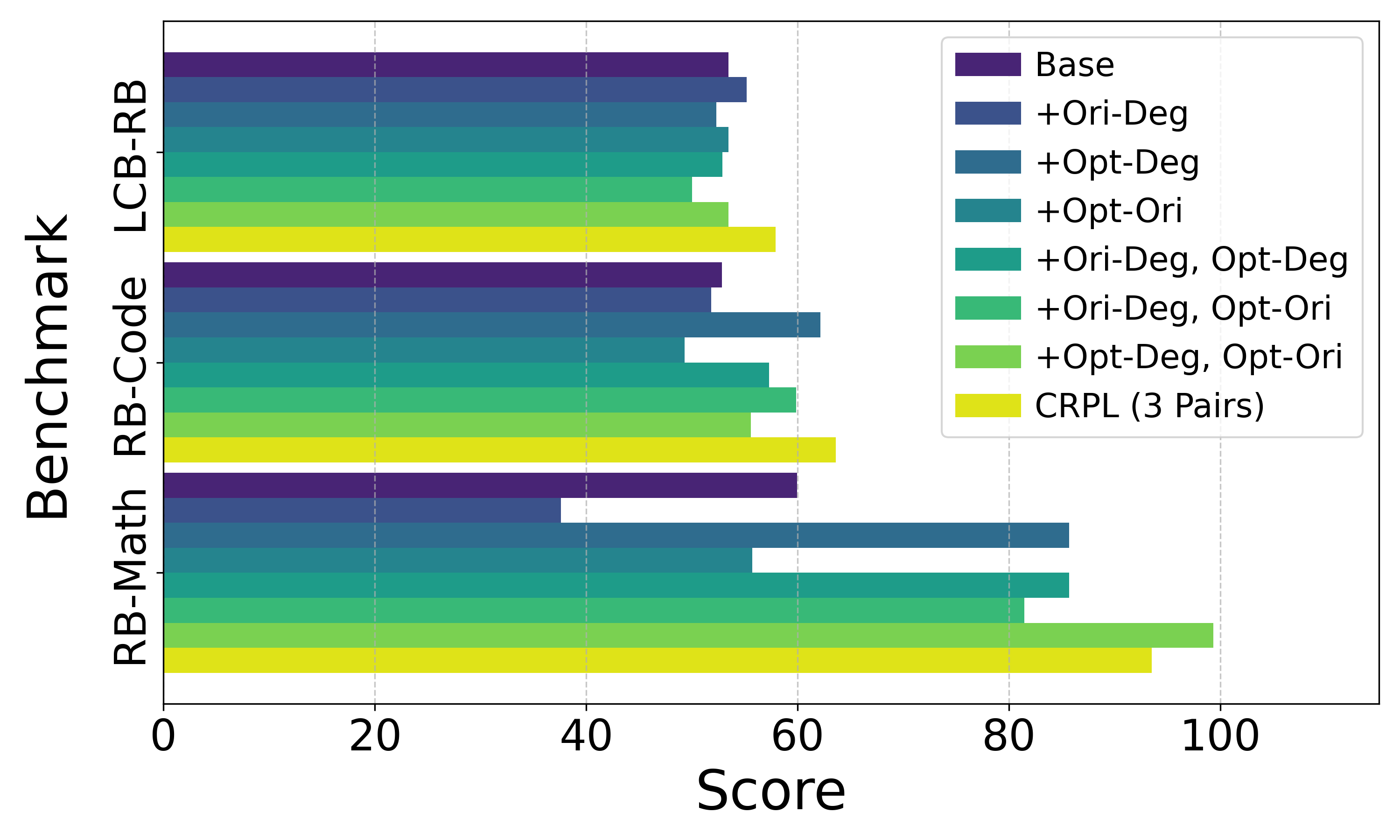}
        \caption{Qwen2.5-Coder-3B.}
        \label{fig:results_3b}
    \end{subfigure}

    \caption{Preference-source ablations across Qwen2.5-Coder backbones.}
    \label{fig:reward_results}
\end{figure}

\textbf{Comparison with Other Reward Models.}
We replace the reward model in \app with Skywork-Reward-Llama-3.1-8B to further validate the effectiveness of our reward model. Due to computational constraints, we use Qwen2.5-Coder-7B-Instruct as the policy model and evaluate performance changes over the first 500 steps on LiveCodeBench. As shown in Figure~\figref{fig:combined-main-results}{fig:reward_comparison}, \app with our \appre-trained reward model demonstrates superior performance. 
This may be because our reward model is trained on fine-grained, reasoning-aware preference pairs, which helps the model capture nuanced differences in reasoning quality. As a result, when used in RL, our reward model yields more precise and informative learning signals for policy optimization than reward models trained on outcome-centric preference data.

\noindent \textbf{The Impact of Reward Hacking.}
To further demonstrate the susceptibility of process-reward RL to reward hacking, we introduce two alternative reward formulations that relax the consistency constraint used in \app. (1) Direct Addition directly adds the process reward to the total reward: $R_i = R^{(i)}_{fmt} + R^{(i)}_{out} + R^{(i)}_{proc}$, (2) Soft-Gated Formulation replaces the hard execution gate with a soft weight based on the execution pass rate: $R_i = R^{(i)}_{fmt} + R^{(i)}_{out} + P^{(i)}_{out} \cdot R^{(i)}_{proc}$, where $P^{(i)}_{out}$ denotes the pass rate of output $o_i$. 
We run controlled experiments using Qwen2.5-Coder-7B-Instruct as the policy and track performance over the first 500 training steps on LiveCodeBench. As shown in Figure~\figref{fig:combined-main-results}{fig:reward_comparison}, both alternative formulations consistently underperform \app, with direct addition yielding the largest drop. These results suggest that reasoning-process rewards computed from incorrect programs are inherently noisy and could be exploited by the policy. Without the strict consistency constraint, the policy may over-optimize $R^{(i)}_{proc}$ even when the resulting code fails execution, ultimately degrading code-generation performance. 
In addition, we visualize the evolution of the reasoning-process reward during training for Direct Addition and ReCode (Figure~\figref{fig:combined-main-results}{fig:reward_hack}). Under Direct Addition, the process reward saturates to near 1.0 within the first 50 steps while downstream performance stagnates, indicating that the policy exploits the unconstrained neural signal. In contrast, under \aapp, the process reward grows gradually and remains moderate, accompanied by steady performance gains, confirming that the consistency gate effectively prevents reward inflation.

\noindent\textbf{Impact of Different Preference Pair Combinations.} We train base models on different combinations with identical experimental settings. As illustrated in Figure~\ref{fig:reward_results}, models achieve optimal performance when trained on all types of preference pairs. This improvement indicates that each pair type contributes useful supervision. We hypothesize that aggregating diverse pair types provides a more comprehensive learning signal, which enables the model to better distinguish reasoning processes. Notably, training solely on Opt–Deg pairs yields the best performance among all single-pair settings, outperforming Opt–Ori and Ori–Deg by 22.2\% on average. This may be because pairs with larger quality separation provide a higher-contrast and less ambiguous preference signal, making it easier for the reward model to learn discriminative features of reasoning-process quality.

\noindent\textbf{Impact of Data Generator Diversity in \appre.}
To investigate whether the reward model overfits to generator-specific stylistic artefacts rather than intrinsic reasoning semantics, we construct a cross-generator variant of \appre by randomly splitting original data into two halves: for one half, we use Llama-3.1-70B-Instruct to synthesize reasoning processes, while the other half is generated by Qwen2.5-Coder-32B-Instruct. We train reward models on the mixed dataset. As shown in Table~\ref{tab:cross}, single-generator training consistently outperforms cross-generator training. For example, with Qwen2.5-Coder-7B as the base model, the single-generator setting improves performance by 11.0\% on average compared to the cross-generator variant. We hypothesize that this gap reflects a signal-to-noise trade-off under a fixed budget: preference pairs produced by a stronger generator tend to provide cleaner pairwise supervision, whereas mixing generators increases stylistic and structural variability of reasoning processes, diluting the preference signal and making it harder for the reward model to learn stable reasoning-process quality distinctions.

\noindent\textbf{Cross-model Generalization.}
To examine whether \aapp generalizes beyond the Qwen2.5 family, we reuse the same CRPL-trained reward model from RQ1 and apply it to Qwen3-4B-Instruct as the policy model. We follow the same training setup as in RQ1, except that training is limited to 400 steps due to computational cost. As shown in Table~\ref{tab:qwen3_generalization}, \aapp yields an 11.1\% relative improvement over outcome-only GRPO on LiveCodeBench. This result suggests that the learned reasoning-process reward transfers across model families.

\begin{table}[t]
\centering
\small
\resizebox{0.9\linewidth}{!}{%
\begin{tabular}{@{}lcccc@{}}
\toprule
\multicolumn{1}{c}{\multirow{2}{*}{Model}} & \multicolumn{3}{c}{LiveCodeBench} & \multirow{2}{*}{Avg} \\ \cmidrule(lr){2-4}
\multicolumn{1}{c}{} & Easy & Medium & Hard &      \\ \midrule
Qwen3-4B-Instruct    & 68.2 & \textbf{32.1}   & 8.3  & 30.7 \\
+GRPO                & \textbf{82.9} & 22.6   & 11.1 & 32.5 \\
+ReCode              & \textbf{82.9} & \textbf{32.1}   & \textbf{12.5} & \textbf{36.1} \\ \bottomrule
\end{tabular}%
}
\caption{Cross-model generalization results on LiveCodeBench.}
\label{tab:qwen3_generalization}
\end{table}

\noindent\textbf{Complementarity with Compiler-based Supervision}
We investigate whether \aapp complements compiler-based process supervision. Following StepCoder~\cite{dou2024stepcoder}, we define a reward based on compilation and test-execution status: $+1$ for passing all unit tests, $-0.3$ for test failures, $-0.6$ for runtime errors, and $-1$ for compilation errors.  We compare three variants trained for 400 steps: (i)~GRPO with the compiler-based reward only, (ii)~\aapp only, and (iii)~\aapp combined with the compiler-based reward.  As shown in Table~\ref{tab:compiler_comparison}, \aapp yields a 5.0\% relative improvement over the compiler-based baseline, while the combined objective achieves the best overall performance. This suggests that the two signals are complementary: compiler-based rewards supervise the correctness of the generated program, whereas \aapp additionally evaluates the semantic quality of the reasoning process and encourages the model to produce more coherent reasoning before code generation.

\begin{table}[t]
\centering
\small
\resizebox{0.9\linewidth}{!}{%
\begin{tabular}{@{}lcccc@{}}
\toprule
\multicolumn{1}{c}{\multirow{2}{*}{Model}} & \multicolumn{3}{c}{LiveCodeBench} & \multirow{2}{*}{Avg} \\ \cmidrule(lr){2-4}
\multicolumn{1}{c}{}   & Easy & Medium & Hard &      \\ \midrule
Qwen2.5-Coder-7B-Instruct   & 56.1 & 3.8    & 6.9  & 18.1 \\
+GRPO (Compiler-based)  & 63.4 & 15.1   & 8.3  & 24.1 \\
+ReCode                & 65.9 & 15.1   & \textbf{9.7}  & 25.3 \\
+ReCode +Compiler-based & \textbf{75.6} & \textbf{17.0}     & 6.9  & \textbf{27.1} \\ \bottomrule
\end{tabular}%
}
\caption{Comparison with compiler-based supervision on LiveCodeBench.}
\label{tab:compiler_comparison}
\end{table}

\noindent\textbf{Generation Efficiency.}
\aapp is also more generation-efficient than outcome-only RL. On LiveCodeBench, \aapp achieves a higher average Pass@1 while generating 23.4\% fewer tokens on average. This suggests that the gains stem from more effective reasoning rather than increased test-time compute. A per-difficulty breakdown is provided in Appendix~\ref{app:efficiency}.

\begin{table}[t]
\centering
\scriptsize

\resizebox{\linewidth}{!}{%
\begin{tabular}{@{}lcccc@{}}
\toprule
\textbf{Model} & \textbf{Size} & \textbf{LCB-RB} & \multicolumn{2}{c}{\textbf{RewardBench}} \\
 & - & - & \textbf{Code} & \textbf{Math} \\ \midrule
Qwen2.5-Coder & 3B & 50.3 & 52.8 & 60.0 \\
+Cross-Gen (50/50) & 3B & 52.9 & 54.6 & 81.7 \\
+Single-Gen & 3B & \textbf{57.9} & \textbf{63.6} & \textbf{93.5} \\ \hdashline
Qwen2.5-Coder & 7B & 53.8 & 43.9 & 65.8 \\
+Cross-Gen (50/50) & 7B & 55.3 & 81.5 & 89.4 \\
+Single-Gen & 7B & \textbf{62.6} & \textbf{88.6} & \textbf{99.8} \\ \bottomrule
\end{tabular}%
}
\caption{Reward model performance under single-generator vs. cross-generator preference data synthesis.}
\label{tab:cross}
\end{table}

\section{Related Work}
\label{sec:related}

\textbf{RL for Code Generation.} Code generation offers a natural verification signal via unit tests~\citep{chencodet,chen2024b4}, enabling outcome-based rewards for RL. Building on this property, a line of work improves LLM coding by leveraging execution feedback. CodeRL~\citep{le2022coderl} applies RL with unit-test feedback, PPOCoder~\citep{shojaee2023execution} enriches rewards by combining unit tests with syntactic and semantic similarity to ground-truth solutions, and DeepSeek-R1~\citep{guo2025deepseek} adopts GRPO using relative pass rates within a group of samples. To address the sparsity of binary signals provided by outcome-based rewards, recent works have shifted towards incorporating fine-grained process supervision into RL. StepCoder~\citep{dou2024stepcoder} uses curriculum learning that progresses from code-completion tasks to full generation, masking unexecuted tokens via test coverage for fine-grained PPO updates.
PRLCoder~\citep{ye2025process} learns a line-level process reward model validated by the compiler and the test cases from mutation and refactoring.
Despite these advances, existing methods remain largely implementation-centric. They primarily evaluate the correctness of generated code rather than the logical soundness of reasoning processes. Motivated by evidence that reasoning quality affects functional correctness, we propose \aapp to explicitly incentivize rigorous reasoning. As prior approaches largely focus on implementation-level refinement, \aapp is potentially complementary to them by improving reasoning-process quality.

\noindent\textbf{Reward Model Evaluation on Reasoning.}
Evaluating the performance of reward models for reasoning tasks typically relies on verifiable problems. 
For example, the code subset of RewardBench~\citep{lambert2024rewardbench} utilizes HumanEvalPack~\citep{muennighoff2023octopack}, a multilingual extension of the HumanEval dataset. 
However, they typically focus only on the correctness of the final output, neglecting the quality of the intermediate reasoning process that produced it.
Additionally, there is a risk of data contamination for benchmarks derived from HumanEval~\cite{jain2025livecodebench}. This can lead to inflated performance metrics that do not reflect true capabilities. 
To address these, we construct \appben sourced from LiveCodeBench~\citep{jain2025livecodebench}, which allows us to reliably evaluate a reward model's discrimination capabilities on the intermediate steps of problem-solving.

\section{Conclusion}
\label{sec:conclusion}

We propose \aapp (\textbf{Re}asoning-\textbf{Re}inforced \textbf{Code} Generation), a novel RL framework designed to align code generation with high-quality reasoning processes. It comprises two components, i.e., \appre for reliable measurement of reasoning quality and \app for safe integration of reasoning-process rewards in RL. Our reward model achieves strong performance on \appben, a new benchmark designed to distinguish between high-quality and flawed reasoning processes. Extensive experiments demonstrate the effectiveness of \aapp, and this paradigm generalizes to the math domain.

\section*{Limitations}

While our work displays many strengths, we highlight three limitations:

\noindent\textbf{Scalability in Long-Context Reasoning Processes.}
We train with a 4K output length due to compute constraints, which limits direct validation on long-horizon settings where reasoning processes can exceed 30K tokens. Scaling \aapp to long contexts requires longer-context training and a long-context generator for the reward model. Future work could explore extending our framework to long-context reasoning processes.

\noindent\textbf{Limited Scale of LCB-RB.}
LCB-RB is derived from LiveCodeBench. To avoid potential data leakage and ensure label reliability, it contains 219 manually verified preference pairs. While this improves the reliability of evaluation, the small scale may limit coverage. Future work could expand reasoning-centric benchmarks using additional problems and further evaluate reward models.

\noindent\textbf{Limited Exploration of Other LLMs}
We observe cross-domain generalization when applying ReCode to Qwen2.5-Math-7B, and cross-architecture transfer to Qwen3-4B-Instruct. However, our evaluation is still limited to models up to 7B parameters, leaving open how \aapp behaves at larger scales. We leave broader scaling studies as future work.

\section*{Ethical considerations}
All the datasets we use to fine-tune LLMs are publicly available and are for research purposes only. Beyond existing datasets, we also create synthetic preference supervision via \appre and curate \appben from LiveCodeBench. These artifacts are derived from programming tasks and model-generated reasoning processes rather than users' personal data. We conduct automated filtering and manual spot-checking to reduce the presence of personally identifying information (PII) and overtly harmful or offensive content in the generated data, and we release artifacts with a research-only intended use under the licenses and access conditions of the original sources. The open- and closed-source LLMs used (e.g., Qwen2.5 series, GPT-4/3.5) have their own training and deployment considerations documented by their creators. Our reward models are used to score reasoning traces, and models trained with \aapp are optimized to generate responses whose reasoning is consistent with the produced implementation. However, we acknowledge that LLMs, including those used in our study, may occasionally produce improper or harmful content. Such outputs are unintended and do not reflect the views or intentions of the authors.
\section*{Acknowledgements}
This research is supported by the National Natural Science Foundation of China (No.92582107) and Zhejiang Provincial Natural Science Foundation of China (No.LZ25F020003).

\bibliography{aaai2026}

\appendix
\section{Appendix}

\subsection{Implementation Details}
\paragraph{Reward Model Setup}
\label{append:reward}
We utilize Qwen2.5-Coder-32B-Instruct~\citep{hui2024qwen2} to generate reasoning process preference data. The reward model is trained with a batch size of 128 and a learning rate of 1e-6 for 2 epochs. We partition the dataset using a 9:1 train-validation split and employ an early stopping strategy based on the validation set.

We compare our approach against several baselines: (1) Original Model: The base model without any additional fine-tuning. (2) SOTA Reward Models:  Current best-performing reward models to validate the competitive performance of our method, including Starling-RM-34B~\citep{starling2023}, EURUS-RM-7B~\citep{yuan2024advancing}, Skywork-Reward-Llama-3.1-8B~\citep{liu2024skywork}, GPT-4-Turbo-2024-04-09~\citep{Achiam2023GPT4TR}, GPT-3.5-Turbo-0125~\citep{brown2020language} and DeepSeek-V3~\cite{liu2024deepseek}. (3) Score-Based reward model: We employ Qwen2.5-Coder-32B-Instruct to score the initial reasoning processes based on the dimensions of reasoning quality, scoring prompt is in Appendix~\ref{append:prompt}, which is also employed by~\citep{fan2025sophiavl}. We then perform SFT on the base model using consistent hyperparameter settings to those with the \appre-based method. 

\subsection{RL Setup}
\label{append:rl}
For BigCodeBench, we use both the full and hard sets with complete configuration. For LiveCodeBench, we utilize the problems from October 2024 to February 2025, in line with prior work~\citep{yang2025qwen3,tian2025not}.  
We compare our approach against several baselines: (1) Original Model: the original model without any additional training.  (2) SOTA Code Models: Current best-performing models on code generation tasks for competitive comparison, including Llama3-Instruct-70B~\citep{dubey2024llama}, Deepseek-Coder-V2-Lite-Instruct~\citep{zhu2024deepseek}, Qwen2.5-Coder-Instruct 14B~\citep{hui2024qwen2}, GPT-4-Turbo-2024-04-09~\citep{Achiam2023GPT4TR}, and GPT-3.5-Turbo-0125~\citep{brown2020language}  (3) SFT on RL Data: The model fine-tuned on the same dataset using identical hyperparameters with SFT. (4) RL without reasoning reward: The model only with outcome and format rewards.

The RL training is conducted using VeRL~\citep{sheng2024hybridflow} on 8 NVIDIA A800 80GB GPUs, with a total batch size of 32 and a maximum output length of 4,096. We employ AdamW optimizer with a constant learning rate of 1e-6 and train for 1,600 steps.
We remove the KL divergence term and adopt token-level policy gradient loss computation and the clip-higher mechanism  with $\varepsilon_{\text{low}} = 0.2, \varepsilon_{\text{high}} = 0.28$ for training stability~\cite{yu2025dapoopensourcellmreinforcement,he2025skywork,wang2025beyond}

\subsection{Prompt used for RL training}
\label{append:promptrl}
\begin{figure}[H]
    \centering
    \includegraphics[width=1\linewidth]{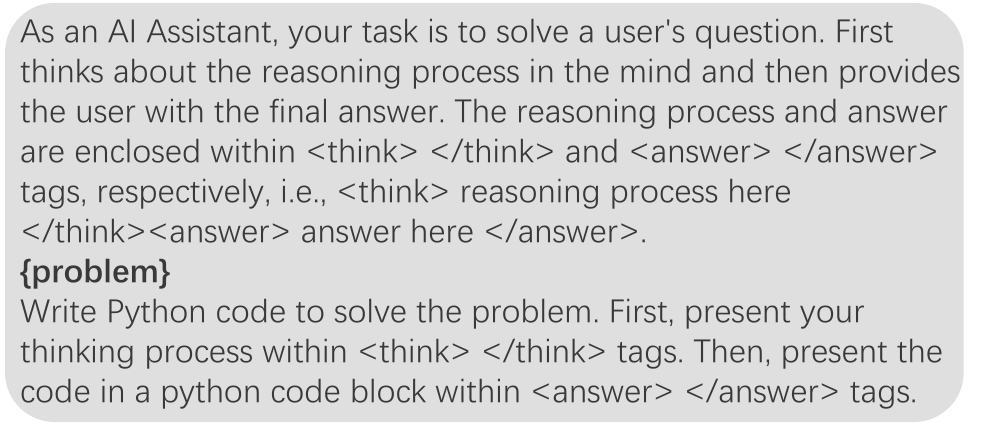}
    \caption{The Prompt used for RL training.}
    \label{fig:rl-prompt}
\end{figure}

\subsection{Synergistic Correlation Between Reasoning Quality and Code Correctness}
\label{append:corr}
To investigate the correlation between reasoning quality and code correctness, we employ a powerful LLM to generate multiple solutions with explicit reasoning processes for coding problems. We then utilize corresponding test cases to categorize the generated code into correct and incorrect implementations. To assess the quality of reasoning processes, we leverage GPT-4o-mini~\citep{Achiam2023GPT4TR} to classify each solution's reasoning into three distinct categories: (1) flawless reasoning with consistent implementation, (2) flawed reasoning with consistent implementation, and (3) inconsistent reasoning and implementation. We exclude the third category from our analysis, as the misalignment between reasoning and implementation introduces confounding factors that would obscure the relationship between reasoning quality and code correctness. This filtering ensures that our study focuses specifically on cases where the implementation faithfully reflects the reasoning process, whether that reasoning is sound or flawed. Consequently, each generated output can be characterized by two attributes: (1) code correctness (correct or incorrect), and (2) reasoning quality (flawless or flawed).

To quantify the association between these two attributes, we perform the chi-square test~\citep{pearson1900x}. Specifically, we utilise Qwen2.5-Coder-32B-Instruct with temperature $T=1.0$ to generate 50 solutions for problems from LiveCodeBench v5, yielding 44,000 candidate outputs in total. After filtering out cases with inconsistent reasoning and implementation, we conduct the statistical test on the remaining aligned set of $N = 33,146$ observations. Table~\ref{tab:contingency} reports the full $2 \times 2$ contingency table. Our analysis yields a highly significant result with $p = 9.3 \times 10^{-15} \ll 0.001$, indicating a strong statistical dependence between reasoning quality and code correctness.

To further validate the reliability of the LLM judge, we randomly sampled 100 cases for independent human verification by two authors. The inter-annotator agreement is Cohen's $\kappa = 0.82$, with disagreements resolved by discussion. The judge achieved a 91\% consistency rate against the human-verified ground truth. Given that this analysis serves as a motivational study to establish broad statistical trends, this level of reliability is sufficient.

\begin{table}[t]
\centering
\small
\begin{tabular}{lrrr}
\toprule
 & Pass & Fail & Total \\
\midrule
Flawless Reasoning & 3,133 & 8,376 & 11,509 \\
Flawed Reasoning   & 5,051 & 16,586 & 21,637 \\
\midrule
Total              & 8,184 & 24,962 & 33,146 \\
\bottomrule
\end{tabular}
\caption{Contingency table.}
\label{tab:contingency}
\end{table}

\subsection{Generation Efficiency Analysis}
\label{app:efficiency}
 
Table~\ref{tab:efficiency} reports the pass rate and average token count on LiveCodeBench by difficulty level. \aapp consistently achieves higher pass rates with fewer generated tokens across all difficulty levels. This indicates that \aapp encourages more concise and targeted reasoning rather than inflating chain-of-thought length.
 
\begin{table}[t]
\centering
\small

\begin{tabular}{@{}lcccc@{}}
\toprule
\multirow{2}{*}{Difficulty} & \multicolumn{2}{c}{GRPO} & \multicolumn{2}{c}{ReCode} \\
\cmidrule(lr){2-3} \cmidrule(lr){4-5}
 & Pass@1 & Avg Tokens & Pass@1 & Avg Tokens \\
\midrule
Easy       & 58.5 & 427.3 & 68.3 & 324.1 \\
Medium     & 15.1 & 568.2 & 20.8 & 441.7 \\
Hard       & 9.7  & 813.6 & 9.7  & 619.8 \\
\bottomrule
\end{tabular}
\caption{Generation efficiency on LiveCodeBench: Pass@1 and average token count across difficulty levels.}
\label{tab:efficiency}
\end{table}

\subsection{Prompt}
\label{append:prompt}
Here, we outline the key prompts utilized in our framework. Figure~\ref{append:gen} shows the prompt for generating the initial reasoning processes. Figures~\ref{append:deg} and~\ref{append:opt} show the prompts for generating degraded and optimized reasoning processes, respectively. Figure~\ref{append:asse} shows the prompt for dual-consistency checking of the reasoning processes. Figure~\ref{append:score} shows the prompt used in the score-based baseline for reward-model training.

\onecolumn	

\begin{figure}[]
\begin{center}
\begin{minipage}{0.8\textwidth}  %
\begin{customprompt}[]{Initial Reasoning Generation Prompt}
\# Task Objective

You are an Expert Problem Solver and Algorithmic Thinker. Your primary goal is to generate a detailed, step-by-step Chain-of-Thought (CoT) that deconstructs and logically solves the given problem. Your output should be the reasoning process itself, not the final solution or code.

\# Input Data

[Problem Statement]

\{problem\_statement\}

\# Requirements for Your Reasoning

1. Deconstruct from First Principles: Begin by dissecting the problem statement. What is the core question? What are the explicit and implicit requirements? What are the inputs, outputs, and constraints? Break the problem down into smaller, more manageable sub-problems.

2. Analyze Examples and Edge Cases: Systematically use the provided examples and test cases to verify your understanding. Explicitly state what each test case teaches you. 

3. Brainstorm and Strategize:

(1) Prioritize Optimal Approaches: Begin by brainstorming efficient strategies. First, explore algorithms and data structures that could lead to an optimal or near-optimal solution (e.g., hash maps, two-pointers, binary search, dynamic programming, greedy algorithms). Do not start by considering the brute-force approach.
    
(2) Select and Justify the Best Strategy: Evaluate the potential efficient approaches you've identified. Choose the most promising one and provide a clear justification for your choice. Analyze its trade-offs in terms of time complexity ($O(n)$), space complexity ($O(n)$), and implementation difficulty. For instance, "A hash map approach offers an optimal $O(n)$ time complexity at the cost of $O(n)$ space, which is an acceptable trade-off here. We will proceed with this strategy."
    
(3) Acknowledge Brute-Force as a Last Resort: Only if you determine that efficient algorithms are not applicable or are excessively complex to implement for the problem at hand, should you then articulate the reasoning for using a brute-force approach. 

4. Develop a Step-by-Step Logical Plan: Based on your chosen strategy, create a clear, logical, and sequential plan.

(1) Mental Walkthrough: "Pre-run" your logic using a specific example. Narrate the state of your variables or data structures at each step of the plan. 
    
(2) Refine and Self-Correct: After the walkthrough, reflect on the plan. Are there any logical gaps? Does it correctly handle all the identified edge cases? Could any step be simplified or made more robust? Acknowledge and address any flaws found during the mental walkthrough.

5. Clarity and Structure: Ensure the entire reasoning process is articulated in a clear, structured manner that is easy for a human to follow. The goal is to illuminate the *how* and *why* of the solution, not just the what.

\# Output Format

Your response must contain ONLY the reasoning process, formatted in Markdown. Do not include any introductory or concluding remarks outside the reasoning block.

\end{customprompt}
\end{minipage}
\end{center}
\caption{Prompt used for initial reasoning generation.}
\label{append:gen}
\end{figure}

\begin{figure}
    \centering
    \begin{minipage}{0.8\textwidth}  %
\begin{customprompt}{Reasoning Degrading Prompt}
\# Task Objective

You are a Red Teaming AI Agent specializing in crafting sophisticated negative training data for advanced reasoning models. Your task is to deliberately introduce a specific, targeted flaw into a `Golden Chain-of-Thought' (CoT). This creates challenging examples that teach other models to identify and avoid logical errors.

\# Input Data

[Problem Statement]

\{question\}

[Golden Chain-of-Thought]

\{golden\_CoT\}

\# Degradation Methods

1. Factually Incorrect Reasoning: Introduce a clear factual error into the logic. For example, misstate a core constraint from the problem, use an incorrect mathematical formula, or misrepresent the time/space complexity of a known algorithm.

2. Irrelevant or Misleading Path: Add steps that are factually correct on their own but are irrelevant to solving the actual problem. This creates a distracting and inefficient reasoning path.

3. Incomplete Reasoning: The reasoning starts correctly but halts before reaching the final step, leaving the logic unfinished and the conclusion unsupported.

4. Logical Gap / Jump: Remove a key intermediate step, making the jump from a premise to a conclusion seem abrupt and unsubstantiated, even if the final conclusion happens to be correct.

5. Chaotic or Acausal Reasoning: Invert the cause-and-effect relationship, or create a sequence of steps that are logically disconnected and do not follow a coherent progression.

\# Execution Steps

1. Identify Methods: Identify one or more `Degradation Methods' from the inputs (e.g., a comma-separated list like ``Logical Gap, Factually Incorrect Reasoning").

2. Analyze \& Plan: Carefully analyze the `Golden CoT'. Strategically plan how to weave all the selected degradation methods into the reasoning. The flaws should be as subtle as realistically possible, modelling a plausible human error.

3. Generate Degraded CoT: Rewrite the CoT to create the flawed `[Degraded CoT]'. This section must contain ONLY the flawed reasoning itself.

4. Generate Explanation: Create a concise `[Explanation of Degradation]'. In this section, you must clearly list each degradation method you used, and for each one, pinpoint exactly how, where, and why you altered the original reasoning.

\# Output Format

Your response MUST be in Markdown format and strictly adhere to the two-part structure below. If multiple degradations are applied, list each one in the explanation.

\texttt{```}markdown

[Degraded Cot]

(Write the Degraded Chain-of-Thought here.)

[Explanation]

(Describe where and how you applied the degradation method(s).)
\end{customprompt}
\end{minipage}
\caption{Prompt used for generating degraded reasoning generation.}
\label{append:deg}
\end{figure}

\begin{figure}[H]
    \centering
    \begin{minipage}{0.8\textwidth}  %
\begin{customprompt}{Reasoning Evolving Prompt}
\# Task Objective

You are an AI Reasoning Optimizer, specializing in refining training data for advanced reasoning models. Your task is to take a Golden Chain-of-Thought (CoT) and apply one or more optimizations to make its logic more rigorous, efficient, and accurate. The goal is to create higher-quality training samples to elevate the performance of advanced reasoning models.

\# Input Data

[Problem Statement]

\{question\}

[Golden Chain-of-Thought]

\{golden\_CoT\}

\# Optimization Methods

1. Factual Verification \& Correction: Identifies and corrects a clear factual error within the reasoning. If no errors are found, this method should not be applied.

2. Focusing Logic: Identifies and removes any redundant steps from the original reasoning. This ensures every step directly contributes to the final goal, making the entire reasoning path more focused.

3. Comprehensive Reasoning: Extends a line of reasoning that may have halted prematurely or omitted final steps. This ensures the logical chain is fully closed and the conclusion is explicitly and robustly supported.

4. Bridging Logical Gaps: Adds necessary intermediate steps between logical nodes that seemed disjointed. This makes the transition from premise to conclusion smoother and more self-evident.

5. Enhancing Logical Flow: Reorganizes reasoning steps to follow a clearer, more intuitive causal or hierarchical order. This ensures the entire thought process is well-structured and flows seamlessly from start to finish.

\# Execution Steps

1. Identify Methods: Based on the `Optimization Methods' above, analyze the input Golden CoT and identify one or more specific methods for application (e.g., a comma-separated list like ``Bridging Logical Gaps, Factual Verification").

2. Analyze \& Plan: Carefully analyze the `Golden CoT'. Formulate a clear strategy for integrating all selected optimization methods into the new reasoning process. The goal of the optimization is to make the reasoning more rigorous, clear, and persuasive.

3. Generate Optimized CoT: Rewrite the CoT to create the `[Optimized CoT]'. This section must contain ONLY the improved reasoning itself.

4. Generate Explanation: Create a concise `[Explanation of Optimization]'. In this section, you must clearly list each optimization method you used and, for each one, pinpoint exactly how, where, and why you improved the original reasoning.

\# Output Format

Your response MUST be in Markdown format and strictly adhere to the two-part structure below. If multiple optimization methods are applied, list each one in the explanation.

\texttt{```}markdown

[Optimized CoT]

(Write the optimized Chain-of-Thought here.)

[Explanation]

(Describe where and how you applied the optimization method(s).)
\end{customprompt}
\end{minipage}
\caption{Prompt used for generating optimized reasoning generation.}
\label{append:opt}
\end{figure}

\begin{figure}
\begin{center}
\begin{minipage}{0.8\textwidth}  %
\begin{customprompt}{Reasoning Flaw Assessment Prompt}
You are a top-tier code reviewer and logical analyst.

Your task is to rigorously analyze a programming solution by evaluating both its thought process (`\textless think\textgreater') and the consistency of its implementation (`\textless answer\textgreater').

Key Analysis Criteria:

1.  Reasoning Soundness: Is the algorithm, logic, and step-by-step plan described in the `\textless think\textgreater` block a correct and robust way to solve the problem? Does this logic have flaws?

2.  Implementation-Thought Consistency: Does the code in the `\textless answer\textgreater' block faithfully implement the logic described in the `\textless think\textgreater' block?

Input Format:

[Problem Description]

\{problem\_description\}

[Solution]

\{solution\_content\}

Your Task:

Strictly adhere to the following two-line output format.

Line 1: Output only `Yes', `No', or `None' based on the following specific logic:

(1) Output `Yes' ONLY if the reasoning in `\textless think\textgreater' has a flaw, AND the code in `\textless answer\textgreater' is a consistent implementation of that flawed reasoning.

(2) Output `No' ONLY if the reasoning in `\textless think\textgreater' is sound, AND the code in `\textless answer\textgreater' is a consistent implementation of that sound reasoning.

(3) Output `None' in all other scenarios. This primarily means any case where the code in `\textless answer\textgreater' is NOT a consistent implementation of the logic in `\textless think\textgreater', regardless of whether the reasoning is sound or flawed.

Line 2: Explain the reasoning for your judgment. Your explanation must address both the soundness of the thought process and its consistency with the final code.
\end{customprompt}
\end{minipage}
\caption{Prompt used for checking the reasoning process.}
\label{append:asse}
\end{center}
\end{figure}

\begin{figure}
\begin{center}
\begin{minipage}{0.8\textwidth}  %
\begin{customprompt}{Reasoning Scoring Prompt}
\# Task Objective

You are an expert evaluator of AI reasoning. I will provide you with a problem and a candidate's chain-of-thought reasoning. Your goal is to judge the quality of this reasoning process and assign it a single score between 0 and 1. Your evaluation must focus on the logical integrity of the process, not merely on whether the final answer is correct.

\# Input Data

[Problem Statement]

\{question\}

[Reasoning Process]

\{reasoning\_to\_evaluate\}

\# Evaluation Criteria

1.  Factual Errors: Does the reasoning introduce incorrect facts, misuse formulas, or misstate constraints from the problem?

2.  Logical Gaps or Jumps: Are there missing steps? Does the conclusion jump from a premise without a clear, logical bridge?

3.  Irrelevant or Misleading Paths: Does the reasoning include steps that, while perhaps factually correct, are irrelevant to solving the problem and create a distracting or inefficient path?

4.  Incomplete Reasoning: Does the reasoning start correctly but stop short of reaching a final, supported conclusion?

5.  Chaotic or Acausal Structure: Is the reasoning jumbled? Does it invert cause-and-effect or present steps in an illogical, disconnected order?

\# Scoring Instructions

Provide a single score from {0, 0.1, 0.2,..., 1.0} based on the reasoning quality.

1.0: Perfectly sound reasoning. Clear, correct, complete, and efficient.

0.7 - 0.9: Minor flaws. Contains small, easily correctable errors or slight inefficiencies.

0.3 - 0.6: Significant flaws. Contains major logical gaps, factual errors, or irrelevant paths that seriously undermine the reasoning.

0.0 - 0.2: Completely flawed. The reasoning is chaotic, nonsensical, or fundamentally wrong from the start.

\#  Output Format

Be strict, you should only output the score without any explanation.
\end{customprompt}
\end{minipage}
\end{center}
\caption{Prompt used in the score-based baseline for reward-moded training.}
\label{append:score}
\end{figure}

\end{document}